\documentclass[twocolumn,10pt,nofootinbib]{revtex4-1}
\pdfoutput=1

\usepackage{gensymb}
\usepackage{array} 
\usepackage{amssymb}
\usepackage{color}
\usepackage{graphicx}
\usepackage{dcolumn}
\usepackage{bbm}
\usepackage{amsmath}
\usepackage{amsfonts}
\usepackage{wasysym}
\usepackage[pdftex,bookmarks,plainpages=false,pdfpagelabels]{hyperref}
\usepackage{slashed}
\usepackage{url}
\usepackage{booktabs}
\usepackage{float}

\renewcommand{\log}[1]{\mathrm{log}\left(#1\right)}
\newcommand{\tr}[1]{\mathrm{tr}\left(#1\right)}

\newcommand{\lmax}{\Lambda_\mathrm{MAX}}

\begin{document}

\begin{mbox}
{\hspace{14.5cm}MPP-2016-168}
\end{mbox}

\title{Fermionic WIMPs and Vacuum Stability in the Scotogenic Model}

\author{Manfred Lindner}\email{lindner@mpi-hd.mpg.de}
\affiliation{Max-Planck-Institut f\"ur Kernphysik,
Saupfercheckweg 1, 69117 Heidelberg, Germany
}

\author{Alexander Merle}\email{amerle@mpp.mpg.de}
\affiliation{Max-Planck-Institut f\"ur Physik,
F\"ohringer Ring 6, 80805 M\"unchen, Germany}

\author{Moritz Platscher}\email{moritz.platscher@mpi-hd.mpg.de}
\affiliation{Max-Planck-Institut f\"ur Kernphysik,
Saupfercheckweg 1, 69117 Heidelberg, Germany
}

\author{Carlos E.~Yaguna}\email{carlos.yaguna@mpi-hd.mpg.de}
\affiliation{Max-Planck-Institut f\"ur Kernphysik,
Saupfercheckweg 1, 69117 Heidelberg, Germany
}

\begin{abstract}
We demonstrate that the condition of vacuum stability severely restricts scenarios with fermionic WIMP dark matter in the scotogenic model. The sizable Yukawa couplings that are required to  satisfy the  dark matter constraint via thermal freeze-out in these scenarios  tend to destabilise the vacuum at scales below that of the heaviest singlet fermion, rendering the model inconsistent from a theoretical point of view.  By means of a scan over the parameter space, we study the impact of these renormalisation group effects on the viable regions of this model. Our analysis  shows that a fraction of more than 90\% of the points compatible with all known experimental constraints -- including neutrino masses, the dark matter density, and lepton flavour violation --  is actually inconsistent. 
\end{abstract}

\maketitle
\thispagestyle{empty}

\section{\label{sec:intro}Introduction}

The scotogenic model~\cite{Ma:2006km} is arguably the simplest radiative scenario that can simultaneously account for dark matter and neutrino masses. In this model the particle content of the Standard Model (SM) is extended by a new scalar doublet ($\eta$) and  three (or two) right-handed singlet fermions ($N_i$). These new fields are further assumed to be odd under a $\mathbb{Z}_2$ symmetry that remains unbroken, while all SM fields are even. In this setup, neutrinos acquire Majorana  masses radiatively at the 1-loop level via diagrams mediated by the new fields, whereas the dark matter can be accounted for by  the lightest $\mathbb{Z}_2$-odd particle -- if it is an electrically neutral scalar or a singlet fermion -- which is rendered stable by the $\mathbb{Z}_2$ symmetry. The phenomenology of this model is extremely rich, covering areas such as   dark matter, neutrino masses, collider searches, and lepton flavour violation, and it has been extensively studied in the literature -- see e.g.~\cite{Kubo:2006yx,Sierra:2008wj,Suematsu:2009ww,Hambye:2009pw,Gelmini:2009xd,Adulpravitchai:2009gi,Adulpravitchai:2009re,Aoki:2010tf,Schmidt:2012yg,Kashiwase:2012xd,Gustafsson:2012aj,Ma:2012if,Kashiwase:2013uy,Klasen:2013jpa,Ho:2013hia,Arhrib:2013ela,Racker:2013lua,Toma:2013zsa,Vicente:2014wga}.

The renormalisation group equations (RGEs) for the scotogenic model have been first computed in~\cite{Bouchand:2012dx} and more recently improved in~\cite{Merle:2015ica}. In relation to these works, it was pointed out  that the RGE corrections could potentially impose strong constraints on the model because they have a tendency to induce the breaking of the $\mathbb{Z}_2$ parity~\cite{Merle:2015gea}. In this paper, we will extend such considerations and investigate further constraints on the model arising from running effects. The main novelty in our analysis is that, unlike previous works, we first impose all low energy constraints -- coming from neutrino masses, precision data, the dark matter density, lepton flavour violating processes, etc.\ -- to obtain, from a random number scan, a large sample of points compatible with \emph{all} known bounds; only then we analyse how the renormalisation group corrections affect the viability of these points.  

Renormalisation group corrections are expected to be particularly important in the case of fermionic WIMP dark matter -- which  will be our focus in the following -- because the Yukawa couplings required to obtain the observed relic density via thermal freeze-out must be sizable in that case.  Such large Yukawa couplings  drive the quartic self-coupling associated with the new doublet toward negative values, destabilising the vacuum at low scales. Interestingly, a study of this effect -- although yielding important consequences -- does not seem to be contained in the literature on the scotogenic model. There do exist several analysis for the inert doublet model (i.e., without singlet fermions): Ref.~\cite{Sokolowska:2011aa} studied how the quartic coupling is affected when radiative effects are included, Refs.~\cite{Khan:2015ipa,Swiezewska:2015paa} went further by demonstrating the impact of vacuum metastability and further consistency constraints on the dark matter sector, and Ref.~\cite{Castillo:2015kba} even investigated the behaviour of the $\mathbb{Z}_2$ symmetry in the inert doublet model. However, all these references focused only on the ``scalar part'' of  the scotogenic model and thus have not revealed the issues lying in its ``fermionic part''.

We specifically determine, for each viable set of points, the highest scale for which such a model remains consistent, denoted by $\lmax$. New physics beyond the scotogenic model should therefore appear \emph{below} $\lmax$, to save the otherwise incompatible setting. Our results indicate that the scale $\lmax$ is always low, often lying below 10~TeV. Many points, in fact, even feature a $\lmax$ smaller than 1~TeV. Notably, we find that in the great majority of cases $\lmax$ is \emph{below} the mass of the heaviest singlet fermion, rendering such otherwise compatible settings inconsistent from a theoretical point of view. As we will see,  only a small fraction of points from our scan can escape this fate. Thus, renormalisation group effects \emph{severely} constrain thermally produced fermionic dark matter within the scotogenic model. 

The remainder of the paper is organised as follows. In the next section we review the scotogenic model and introduce our notation. Section~\ref{sec:constraints} presents the most relevant theoretical and experimental constraints that must be satisfied. Our main results are laid out in section~\ref{sec:results}. We discuss some implications of our results in section~\ref{sec:consequences} and finally draw our conclusions in section~\ref{sec:conclusions}.



\section{The model}

The scotogenic model is a simple extension of the SM by a second scalar doublet $\eta$ and (usually) three generations of right-handed singlet fermions $N_i$~\cite{Ma:2006km}. All new fields are assumed to be odd under a discrete global $\mathbb{Z}_2$ parity. The Lagrangian of this model includes the following terms
\begin{equation}
  \mathcal{L} = \mathcal{L}_\mathrm{SM} - \frac{1}{2} M_{i} \overline{N_i} N_i^\mathcal{C} + h_{ij}  \overline{N_i}\, \widetilde{\eta}^\dag\, \ell^j + \mathrm{h.c.} + V,
\end{equation}
where $M_i$ are the Majorana masses of the singlet fermions while $h_{ij}$ is a new matrix of Yukawa couplings, which we take to be real. The scalar potential, $V$, can be explicitly written as:
\begin{align}
    V &= m_H^2 \phi^\dag \phi + m_\eta^2 \eta^\dag \eta + \frac{\lambda_1}{2} \left(\phi^\dag \phi\right)^2 \nonumber\\
      & + \frac{\lambda_2}{2} \left(\eta^\dag \eta\right)^2 + \lambda_3 \left(\phi^\dag \phi\right)\left(\eta^\dag \eta\right) \\
      & + \lambda_4 \left(\phi^\dag \eta\right)\left(\eta^\dag \phi\right) + \left[ \frac{\lambda_5}{2} \left(\eta^\dag \phi \right)^2 + \mathrm{h.c.} \right]. \nonumber
\end{align}
Upon electroweak symmetry breaking (EWSB), this potential yields four  physical scalar particles, denoted by $h,\,\eta^\pm,\,\eta_R,$ and $\eta_I$, where $h$ is the SM Higgs boson observed at the LHC with a mass of about 125 GeV. Their squared masses are given by, respectively,
\begin{subequations}\label{eq:scalarMasses}
\begin{align} 
    m_h^2 &= 2 \lambda_1 v^2 = - 2 m_H^2,\\
    m_\pm^2 &= m_\eta^2 + v^2 \lambda_3,\\
    m_R^2 &= m_\eta^2 + v^2 \left(\lambda_3 + \lambda_4 + \lambda_5\right),\\
    m_I^2 &= m_\eta^2 + v^2 \left(\lambda_3 + \lambda_4 - \lambda_5\right).
\end{align}
\end{subequations}
With these ingredients, the loop-induced active neutrino mass matrix can be calculated as~\cite{Ma:2006km}:\footnote{We have corrected for a missing factor of $1/2$ that was not contained in the original paper, see the first version of Ref.~\cite{Merle:2015ica} for details.}
\begin{equation}\label{eq:neutrinoMass}
\small
  {m_\nu}_{ij} = \sum_{k=1}^n\frac{M_k h_{ki} h_{kj}}{32 \pi^2} \left\lbrace \frac{m_R^2}{m_R^2-M_k^2} \log{\frac{m_R^2}{M_k^2}} - 
  (R \mapsto I)\right\rbrace.
\end{equation}
Note that, in the limit where $m_R = m_I$, one obtains ${m_\nu}_{ij}=0$. Closer inspection of the expression above reveals that in this case $\lambda_5=0$  [cf.\ Eqs.~\eqref{eq:scalarMasses}], and the Lagrangian has a global $U(1)$ lepton-number-type symmetry which forbids neutrino masses. Consequently, $\lambda_5$ can be small without fine-tuning~\cite{'tHooft:1979bh}, as shown in~\cite{Bouchand:2012dx}. Similar arguments can be given for $h_{ij}$ and $M_k$~\cite{Merle:2015ica}.

The $\mathbb{Z}_2$ symmetry of the scotogenic model ensures that the lightest odd particle is stable and therefore a dark matter candidate, if electrically neutral. Hence, depending on the choice of parameters, we have two possible dark matter candidates: the lightest neutral scalar or the lightest singlet fermion. Throughout this paper we will  be concerned with the region of parameter space where  the lightest singlet fermion, denoted by $N_1$,  accounts for WIMP dark matter.

The renormalisation group equations for this model were derived and studied previously in~\cite{Bouchand:2012dx,Merle:2015ica}. Since they do play a central role in our study, we reproduce them in the Appendix. 

\section{\label{sec:constraints}Constraints}

\subsection{Theoretical constraints}

To ensure that the scalar potential of the scotogenic model is bounded from below and that the vacuum is stable, the following conditions must hold~\cite{Branco:2011,Maniatis:2006,Klimenko:1984}:
\begin{equation}\label{eq:BoundedFromBelow}
\begin{gathered}
 \lambda_1 > 0,\
 \lambda_2 > 0,\
 \lambda_3 > - \sqrt{\lambda_1 \lambda_2},\\
 \lambda_3 + \lambda_4 - |\lambda_5|>- \sqrt{\lambda_1 \lambda_2}.
\end{gathered}
\end{equation}


We also require the Yukawa and scalar couplings to be perturbative, so that our tree-level and one-loop results can be trusted.  For definiteness we impose  $|h_{ij}|^2,|\lambda_{2,3,4,5}| \lesssim 4\pi$.

\subsection{Experimental constraints}
Regarding dark matter, we consider the standard thermal freeze-out scenario to obtain the $N_1$ relic density, i.e., it is taken to be a Weakly Interacting Massive Particle (WIMP). Hence, the dark matter density is assumed to be the result of a freeze-out process, driven by dark matter self-annihilations in the early Universe. Note that $N_1$'s annihilate into leptonic final states via $t$-channel processes mediated by the $\mathbb{Z}_2$-odd scalars, so that the dark matter constraint restricts not only the $N_1$ mass but also  the sizes of the new Yukawa couplings and of the masses of the scalars. All the viable points we are going to consider feature a dark matter relic density, calculated numerically with {\tt micrOMEGAs}, compatible with the Planck determination~\cite{Ade:2015xua}, $\Omega_{N_1} h^2\approx 0.12$. Current bounds from direct or indirect dark matter detection experiments are not relevant for this setup~\cite{Ibarra:2016dlb}.  

The constraints from neutrino masses and mixing angles can be taken into account easily by using a modified version of the Casas-Ibarra parametrisation~\cite{Casas:2001sr}, as explained e.g.\ in~\cite{Toma:2013zsa}. We require compatibility with current neutrino data at $3\sigma$ according to~\cite{Capozzi:2013csa}. When combined with the dark matter constraint, which requires sizable Yukawa couplings, the neutrino data enforces a tiny value for $\lambda_5$. In this setup, neutrino masses are thus small because of $\lambda_5\ll 1$. Note that no further assumptions are made on the structure of the Yukawa matrices.

Lepton flavour violating processes usually set very strong constraints on this scenario, as emphasised in~\cite{Adulpravitchai:2009gi,Vicente:2014wga}. The rates of these processes were calculated for the scotogenic model in~\cite{Toma:2013zsa}, where the full analytical expressions can be found. In our analysis, we impose the current experimental limits on all the relevant processes of this type: $\mathrm{BR}(\mu\to e\gamma)<5.7\times 10^{-13}$~\cite{Adam:2013mnn}, BR$(\mu\to 3e)<1.0\times 10^{-12}$~\cite{Bellgardt:1987du}, CR$(\mu\mbox{-}e, \mathrm{Ti})<4.3\times 10^{-12}$~\cite{Dohmen:1993mp}, BR$(\tau\to \mu\gamma)<4.4\times 10^{-8}$~\cite{Aubert:2009ag} and BR$(\tau\to e\gamma)<4.4\times 10^{-8}$~\cite{Aubert:2009ag}.

For completeness, we have also taken into account the bounds on the scalar masses coming from electroweak precision data~\cite{Baak:2011ze,Goudelis:2013uca} and from collider searches, namely Higgs decays and di-lepton searches~\cite{Pierce:2007ut,Lundstrom:2008ai,Belanger:2015kga}. However, these do not present the relevant constraints for the parameter space of the model.

\section{\label{sec:results}Results}

In this section we present our main results. First, we randomly scan the parameter space of this model to obtain a large sample of points compatible with \emph{all} theoretical and experimental constraints at low energies. Then, we numerically demonstrate that renormalisation group effects strongly affect the viability of these settings, rendering many of the points found inconsistent from a theoretical point of view. Finally, we  show  that this result can be understood analytically from the RGEs.

\subsection{The viable parameter space}
\begin{figure*}[t]
  \includegraphics[width=.45\textwidth]{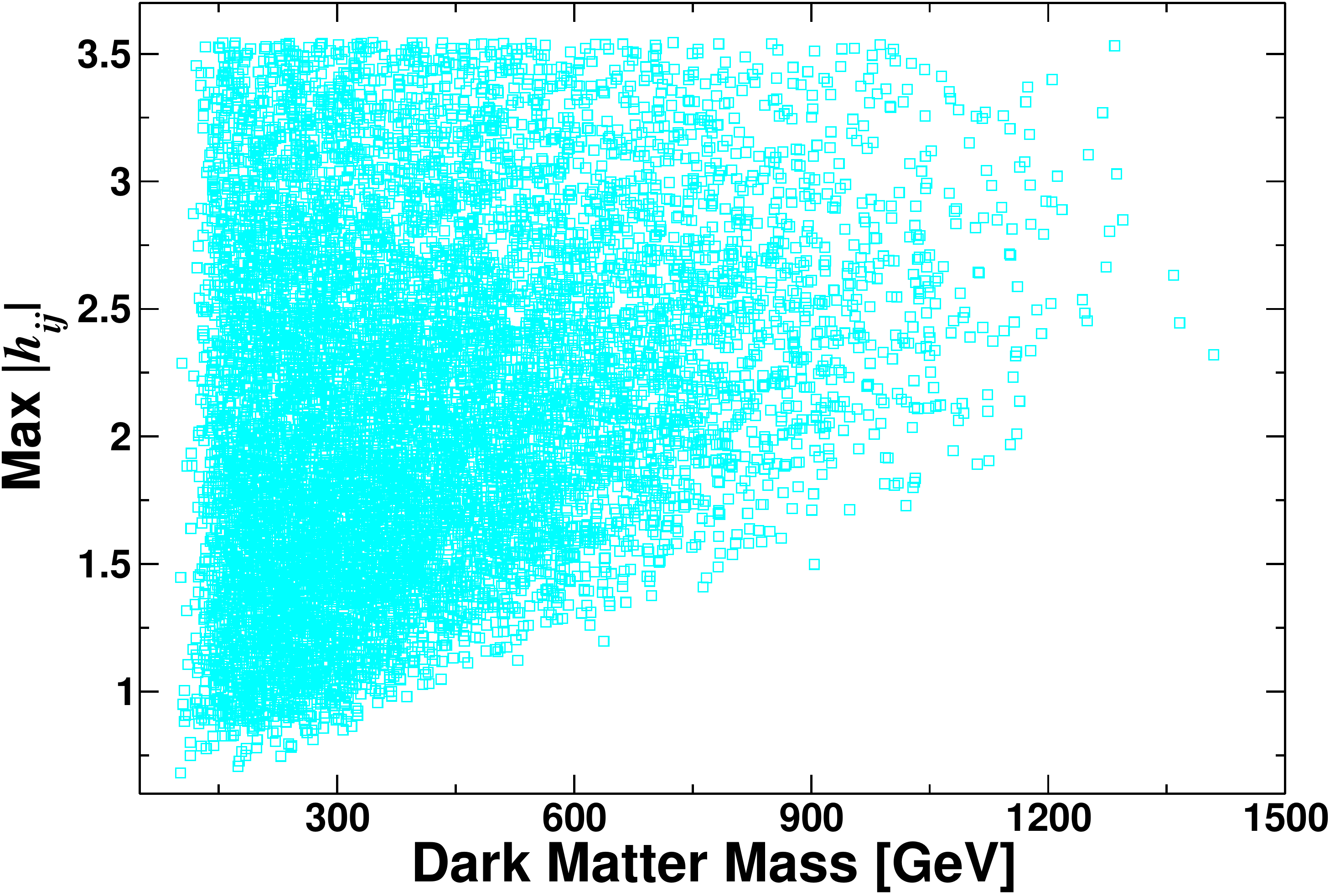}
  \hfill
  \includegraphics[width=.45\textwidth]{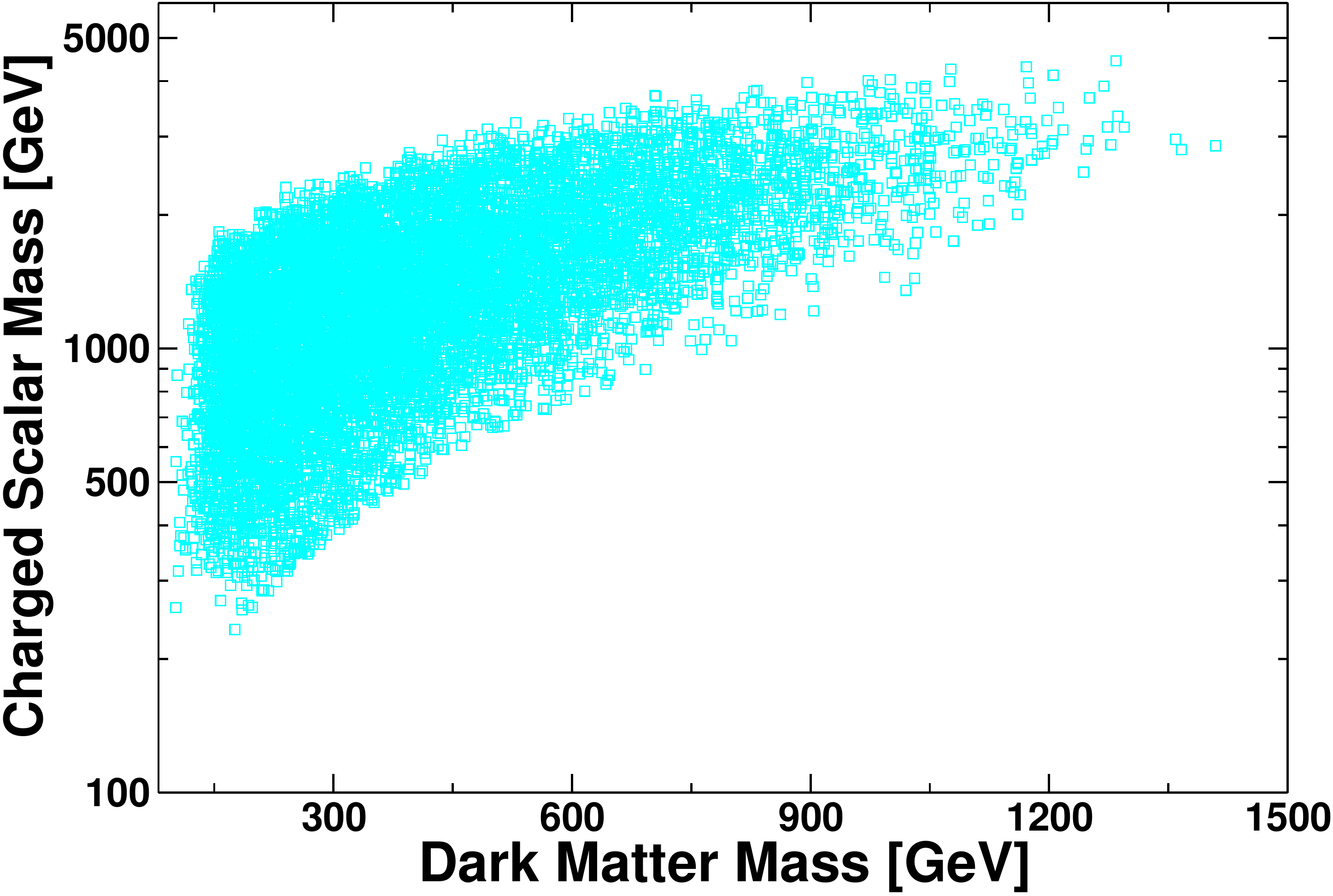}
  \caption{\label{fig:pamspace} Our set of viable points projected onto two different planes. \emph{Left}: The dark matter mass versus the maximum size of the $h_{ij}$ couplings. \emph{Right}: The dark matter mass versus the charged scalar mass.}
\end{figure*}

The scotogenic model introduces $17$  new parameters as follows: $3$ masses for the singlet fermions ($M_i$);  $5$ parameters in the scalar sector, which can be taken to be the scalar couplings $\lambda_{2\ldots 5}$ and the mass of the charged scalar ($m_\pm$); and $9$ new Yukawa couplings ($h_{ij}$ taken as real parameters). These $17$ parameters are, however, not entirely free, as discussed in the previous section. The constraints from neutrino masses and mixing angles, for example, allow us to write the  $9$ Yukawas in terms of just $3$ angles (denoted by $r_i$), eliminating $6$ of them. The remaining  set of $11$ free parameters determines what we call the parameter space of this model. 

We randomly scanned this parameter space within the following ranges:
\begin{align}
100~\mbox{GeV}<&M_i<10~\mbox{TeV}\,,\\
&m_\pm<10~\mbox{TeV}\,,\\
&\left|\lambda_{2,3,4}\right|< 4\pi\,,\\
10^{-8}<&\left|\lambda_5\right|<10^{-12}\,,\\
0<&r_i<2\pi\,,
\end{align}
and imposed all the theoretical and experimental constraints mentioned in the previous section. Finally, we obtained a sample of $10^4$ points compatible with \emph{all} the known phenomenological bounds. This sample represents the viable parameter space for fermion dark matter in the scotogenic  model. 

The viable parameter space is illustrated in FIG.~\ref{fig:pamspace}, where it has been projected onto the planes ($M_{1},\mathrm{max}|h_{ij}|$) on the left panel and  ($M_{1},m_\pm$) on the right panel. Notice that, in particular, the dark matter mass does never exceed $1.5$ TeV in our sample. The right panel shows  that the mass of the charged scalar instead lies below $5$ TeV. From the left panel, we see that some Yukawa couplings are always sizable, an event that can be explained by the WIMP dark matter relic density constraint and by the fact that the annihilation cross section for Majorana fermions is velocity-suppressed. This observation has very important implications regarding renormalisation group effects, as we will show below: the large Yukawa couplings will be the main driving force behind the strong running of the scalar potential parameters.

\subsection{Numerical analysis}
Now that we have imposed all relevant phenomenological bounds and obtained the viable parameter space for fermion dark matter in the scotogenic model, we would like  to determine how the renormalisation group evolution  affects the  consistency of these viable points. This evolution may lead to the violation, at higher scales, of the theoretical constraints mentioned in the previous section. Specifically, we could find that one of the two following outcomes is realised at scales above $M_Z$: 
\begin{enumerate}
\item The vacuum is unbounded or unstable.

\item Some couplings are non-perturbative.\footnote{Note that this latter requirement is not a constraint coming from physics but rather a technical constraint stemming from the fact that the Feynman diagram method is basically invalidated for non-perturbative couplings.}
\end{enumerate}

In our analysis, we follow the renormalisation group evolution of each viable point from the weak scale\footnote{For $\lambda_1$ the input scale is chosen to be $\mu=M_Z$, while the remaining scalar parameters are fixed at the inert scalar threshold $\mu=m_\eta$.} up to the scale $\lmax$, at which  one of these conditions is satisfied. Only up to the scale $\lmax$, therefore, can the scotogenic model  provide a consistent and reliable description of Nature. In other words, \emph{further} new physics beyond the scotogenic model should appear below $\lmax$ -- or we have to completely discard the scenario. 

\begin{figure}[tb]
  \includegraphics[width=.45\textwidth]{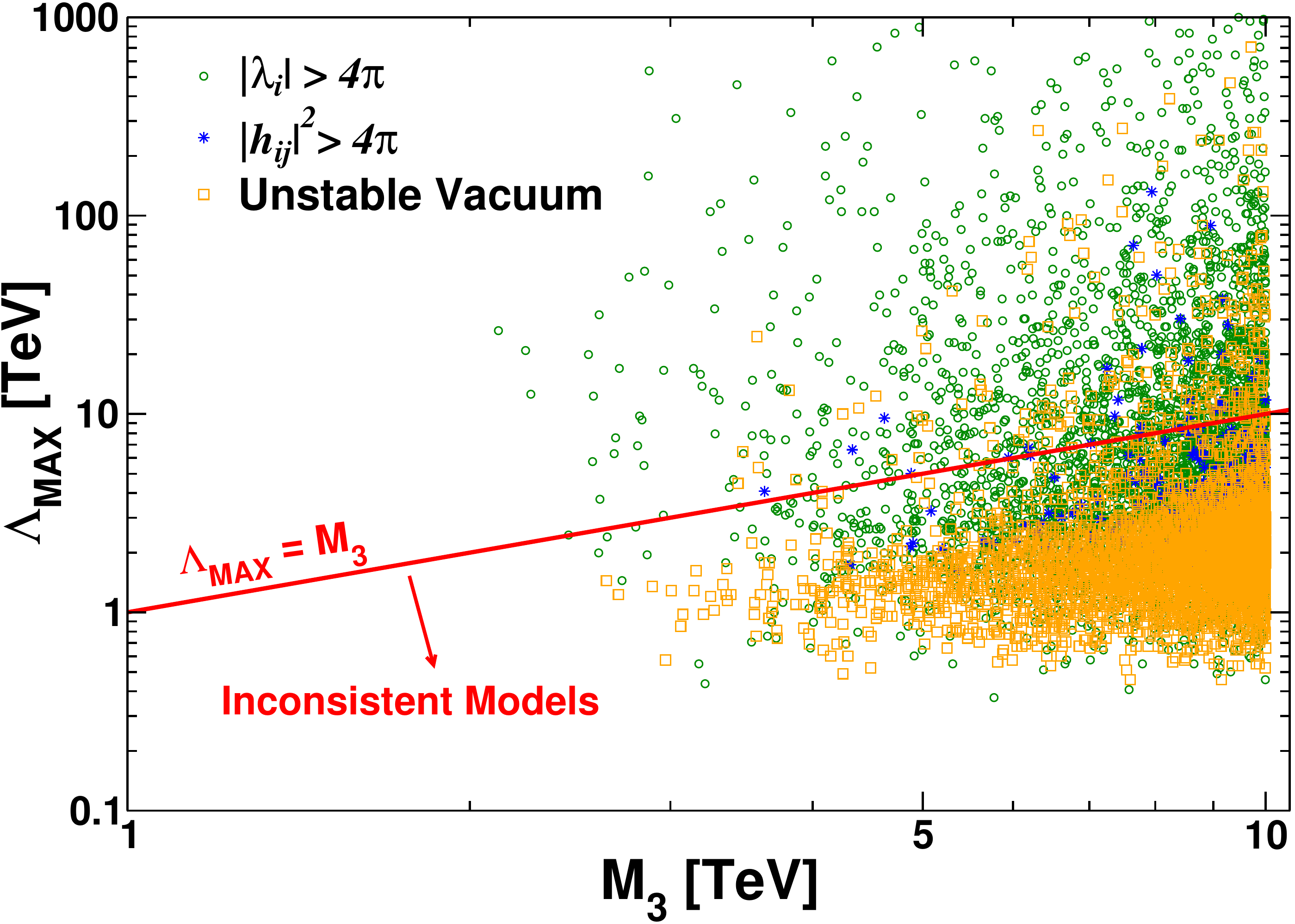}
  \caption{\label{fig:inconsistency} Scale of inconsistency $\Lambda_\mathrm{MAX}$ as a function of the heaviest mass scale in the model. All models below the red line are inconsistent. The color coding indicates the reason for the inconsistency.}
\end{figure}

FIG.~\ref{fig:inconsistency} displays $\lmax$ for our sample of viable models. As abscissa we have used the mass of the heaviest singlet fermion, $M_3$, which happens to be the highest mass scale in this model. Notice that  $\lmax$ is never very high, often lying below $10$ TeV  and in many cases reaching values below $1$ TeV. The color code in this figure denotes the criterion that fails at $\lmax$: Vacuum stability (orange),  perturbativity of Yukawa couplings (blue), or perturbativity of the scalar couplings (green). We found that they account, respectively, for about $50\%$, $3\%$, and $47\%$ of the viable points in our sample.  Notice, from the figure, that the vacuum stability condition tends to be violated at low scales.

\begin{figure*}[tb]
  \includegraphics[width=.45\textwidth]{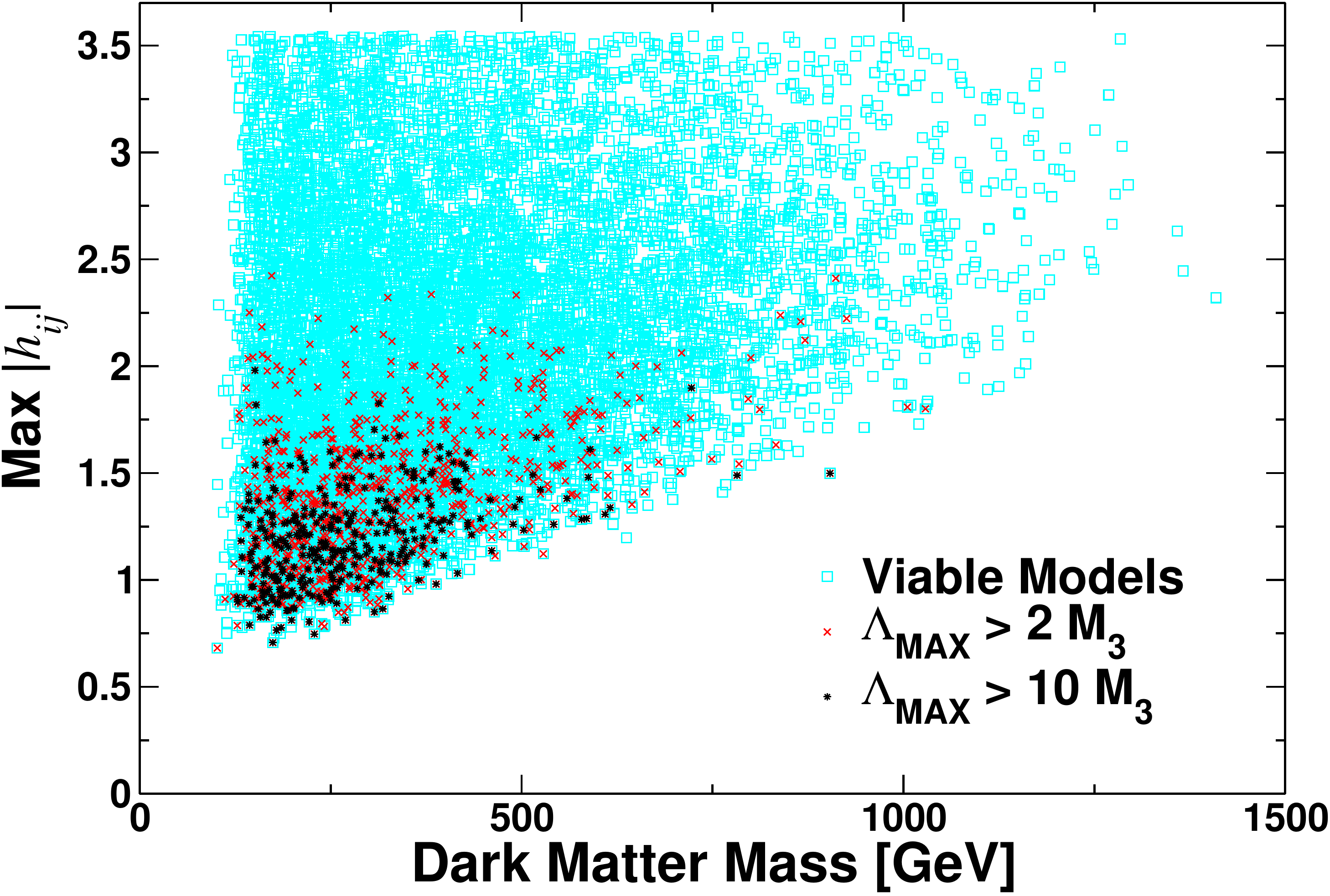}
  \hfill
  \includegraphics[width=.45\textwidth]{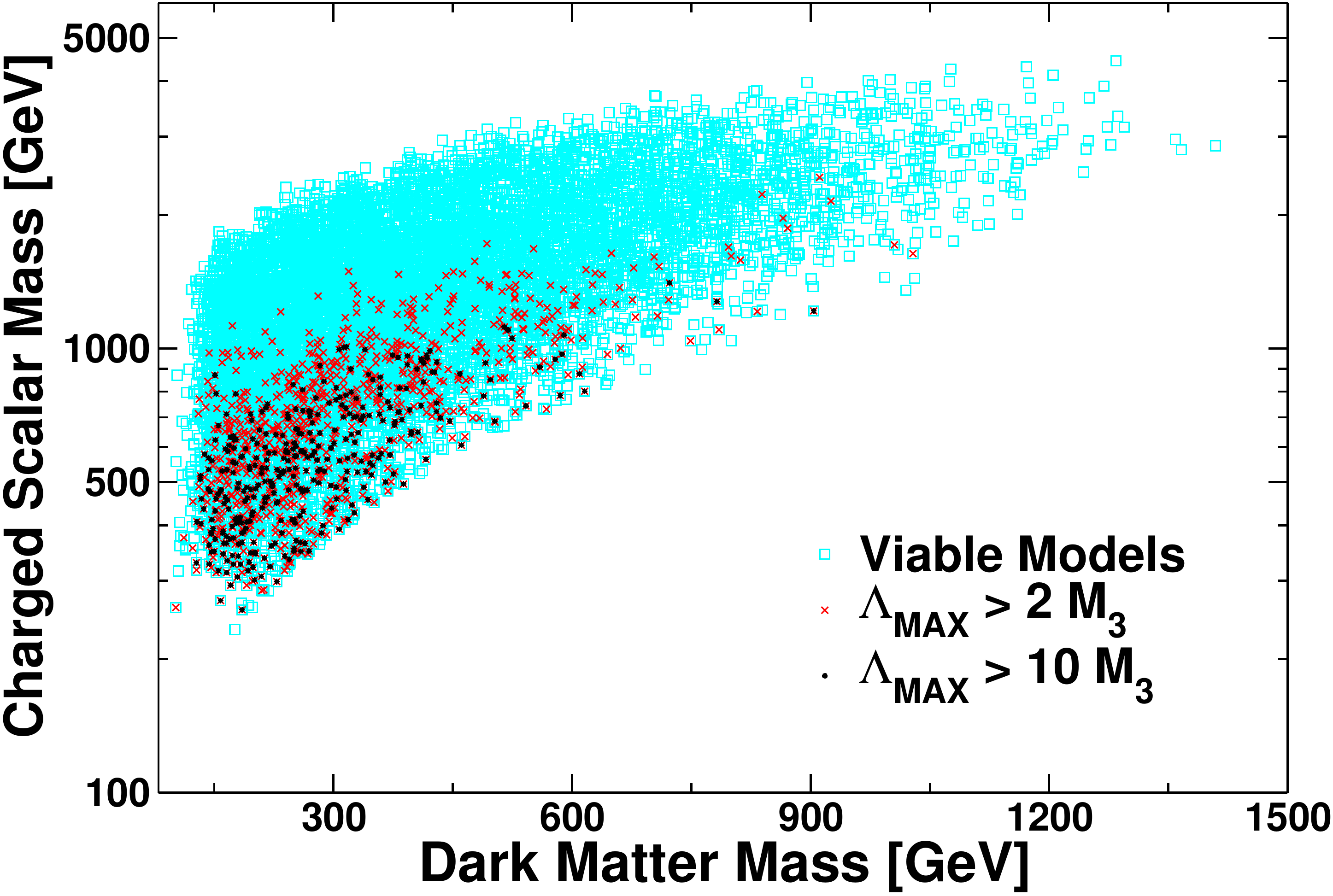}
  \caption{ \label{fig:viable} The modification of the parameter space once renormalisation group effects are taken into account. Requiring consistency of the model up to a scale $10\,M_3$ ($2\,M_3$) leaves only about $3\%$ ($10\%$) of the viable parameter points.  \emph{Left}: The dark matter mass versus the maximum value of the $h_{ij}$ couplings. \emph{Right}: The dark matter mass versus the charged scalar mass.}
\end{figure*}

The red line in FIG.~\ref{fig:inconsistency} corresponds to $\lmax=M_3$. Any parameter point below that line is inconsistent from a theoretical point of view, as new physics would be required below a physical mass scale intrinsic  to the model.  As can be seen in the figure, the large majority of otherwise phenomenologically viable points lie below that line and are, therefore,  actually inconsistent. This fact is the main result of this paper. Thermally produced fermionic dark matter in the scotogenic model is thus severely restricted by renormalisation group effects.

To illustrate the regions of the parameter space that remain  consistent once renormalisation group effects are taken into account, we have superimposed on the viable parameter points (cyan squares)  those satisfying $\lmax>2M_3$ (red crosses) and $\lmax>10M_3$ (black points) -- see FIG~\ref{fig:viable}. As seen clearly in these figures, the number of consistent models is greatly reduced. It amounts to $10\%$ of the models for $\lmax>2M_3$ and $3\%$ for $\lmax>10M_3$. Notice that, in particular, these few viable models tend to feature comparatively small values of the Yukawa couplings, a behaviour that can be understood analytically.



\subsection{Analytical estimates}

In this model, vacuum stability is usually violated when $\lambda_2$ becomes negative, an effect due to the term $\beta_{\lambda_2} \sim-4 \tr{h^\dag h\, h^\dag h}$, cf.\ Eq.~\eqref{eq:RGlambda2}.   
If this term dominates the RGE for $\lambda_2$, we can find a simple estimate for the scale where $\lambda_2 = 0$, i.e., the scale where Eq.~\eqref{eq:BoundedFromBelow} is violated. Neglecting the running of the Yukawa couplings, one obtains:
\begin{equation}\label{eq:Lambda_la2}
  \log{\frac{\Lambda_{\lambda_2}}{\mu_0}} = \left. \frac{4\pi^2 \lambda_2}{\tr{h^\dag h\, h^\dag h}}\right|_{\mu=\mu_0}.
\end{equation}
Note that the remaining quartic couplings do not contain such large terms as the appearance of $h$ is always accompanied by the (tiny) charged lepton Yukawa couplings [cf. Eqs.~\eqref{eq:lambdaRGEs}]. Thus, for very large Yukawa couplings, the conditions~\eqref{eq:BoundedFromBelow} are violated mostly through the running of $\lambda_2$. In some cases this means that, e.g., the condition $\lambda_3+\lambda_4- \left|\lambda_5\right| > - \sqrt{\lambda_1\lambda_2}$ is violated at a scale \emph{below} $\Lambda_{\lambda_2}$ (such that $\lambda_2>0$). However, the large Yukawa contribution will eventually drive $\lambda_2$ to negative values. Thus, we do not estimate this potentially lower inconsistency scale since the setting would be excluded anyway, just at a slightly higher scale.

\begin{figure}[tb]
  \includegraphics[width=.45\textwidth]{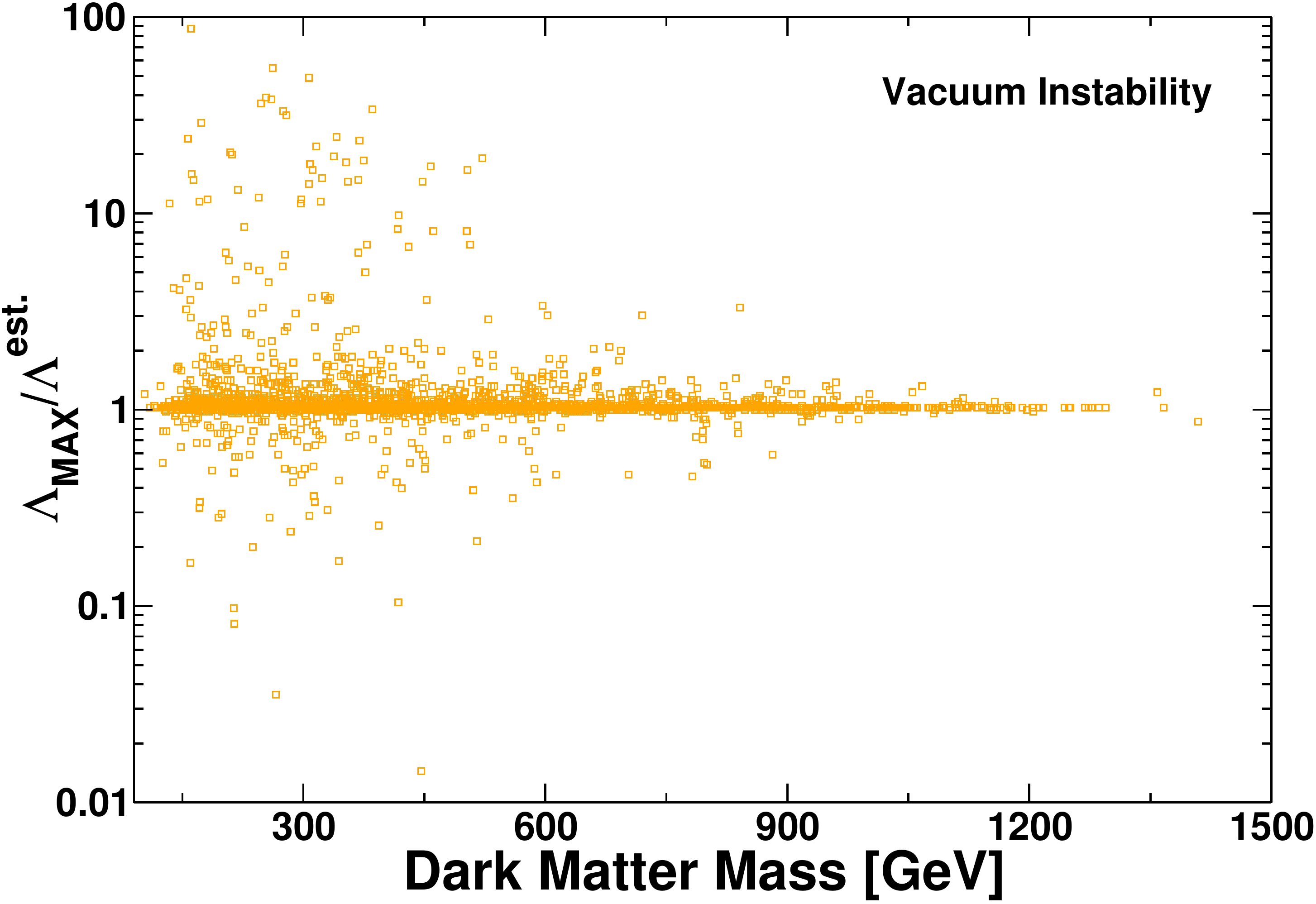}
  \caption{\label{fig:est_vs_num} Ratio of estimated and numerically determined scale of inconsistency for parameter points that violate the vacuum stability condition.}
\end{figure}

Similarly, the scalar couplings may be driven into non-perturbative magnitudes either by the running or by choice of the input values. A simple estimate can be found in this case, assuming that it is the coupling itself that dominates the RGE. In this case the RGE takes the form $\beta_\lambda \sim b \lambda^2$, where $b$ depends on which coupling is considered [cf.\ Eqs.~(\ref{eq:RGlambda1}--\ref{eq:RGlambda5})]. The exact solution to this simplified RGE yields for the scale where $|\lambda_i| = 4\pi$:
\begin{equation}
  \log{\frac{\Lambda_{4\pi}}{\mu_0}} = \left.4\pi\, \mathrm{sgn}(\lambda_i) \frac{4 \pi - \lambda_i}{b\, \lambda_i}\right|_{\mu=\mu_0}.
\end{equation}

To assess the quality of these analytical estimates, we displayed in FIG.~\ref{fig:est_vs_num} the ratio between  $\lmax$ and $\Lambda_{\lambda_2}$ for models that  violate vacuum stability. In most cases, the estimate gives  an accurate estimate of the scale $\lmax$. Thus, using the above equations, it is possible to estimate $\lmax$ directly from the low energy data.

In addition, Eq.~\eqref{eq:Lambda_la2} also indicates that  the violation of vacuum stability is  closely tied to the magnitude of the Yukawa couplings, which must necessarily be sizable to satisfy the dark matter constraint with thermally produced fermionic WIMP dark matter. 



\section{\label{sec:consequences}Discussion}

As we have seen in the previous section, WIMP-like fermionic dark matter in the scotogenic model is tightly constrained by vacuum stability. 
Although this result was based on a random scan of the parameter space, it does not strongly depend on the specific details of the scan. It is, at the end, a phenomenological requirement -- namely the dark matter constraint -- that  forces the Yukawas to be sizable, driving $\lambda_2$ towards negative values.   In fact, we also did other scans with different ranges for the free parameters, finding  results qualitatively similar to those shown in FIG. \ref{fig:inconsistency}. In all cases, a large fraction of points becomes inconsistent once renormalisation group effects are taken into account. When these numerical checks are combined with the analytical insights from  the previous section, it becomes clear that the violation of vacuum stability is actually  an intrinsic and important feature of the scotogenic model with fermionic WIMP dark matter.

In contrast, the violation of  the perturbativity criterion by the scalar couplings mostly reflect the initial conditions, as the phenomenology does not require large values for them.  We explicitly checked that, for instance, such points can be largely eliminated without modifying the rest of the parameter space in a significant way, simply by requiring smaller scalar couplings at the input scale (e.g.\  $|\lambda_i|<1$ at $m_\eta$).

To avoid problems with vacuum stability, we need to find ways of explaining the dark matter that do not require large Yukawa couplings. Several possibilities may be pursued. Within the freeze-out paradigm, coannihilations between the singlet fermions and the scalars could be used to explain the relic density. These coannihilation effects have already been shown to lead to smaller Yukawa couplings~\cite{Vicente:2014wga}, but they require an unexplained degeneracy between the fermions and the scalars. Another interesting possibility is to produce singlet fermions with very small Yukawa couplings via freeze-in~\cite{Hall:2009bx}, as put forward in~\cite{Molinaro:2014lfa}. In that case, the Yukawas associated with the lightest singlet fermion -- the dark matter particle -- must, however, be really tiny (to prevent thermalisation in the early Universe), lying between $10^{-6}$ and $10^{-12}$ for dark matter masses between $1$~keV and $100$~GeV, respectively; the remaining Yukawas can naturally be small so as to explain neutrino masses. Scalar dark matter provides another straightforward way of avoiding this problem. Since the relic density of scalar dark matter is mostly determined by the gauge interactions, the Yukawa couplings can be taken to be  small without problems. Finally, one could also consider extensions of the scotogenic model, as recently analysed e.g. in~\cite{Merle:2016scw,Ahriche:2016cio}. And, if either the particle content or the gauge group is extended, one could have a setting where the singlet fermions only have very feeble interactions; this could lead to scenarios featuring light fermion dark matter produced via decays (see, e.g.,~\cite{Kusenko:2009up,Petraki:2007gq,Merle:2013wta,Klasen:2013ypa,Merle:2014xpa,Merle:2015oja,Shakya:2015xnx}) and diluted thermal production~\cite{Bezrukov:2009th,King:2012wg,Nemevsek:2012cd,Patwardhan:2015kga}, respectively. This is similar but not identical to sterile neutrino dark matter, due to the absence of active-sterile mixing in the scotogenic model.

\section{\label{sec:conclusions}Conclusions}

We have demonstrated that the vacuum stability condition severely restricts the  viability of thermally produced fermionic dark matter in the scotogenic model. The reason for these effects being so important in this scenario is that large Yukawa couplings are required to satisfy the relic density constraint. These large Yukawas tend to  destabilise the vacuum  at scales below that of the heaviest singlet fermion, rendering the scotogenic model inconsistent from a theoretical point of view in a significant part of the parameter space. We investigated these effects in some detail, both  numerically and analytically. By means of a scan over the parameter space, we studied the impact of renormalisation group effects on the viable regions of the model. We showed that, specifically, the vast majority of  points compatible with all known experimental constraints -- including neutrino masses, the dark matter density, and lepton flavour violation --  are  actually inconsistent. Moreover, the violation of vacuum stability, driven by the large Yukawas, was identified as the primary factor that sets the inconsistency scale for most viable points. In addition, we  found reliable analytical estimates for the inconsistency scale, and we  briefly explored ways out of this problem.

\begin{acknowledgments}
MP acknowledges support by the IMPRS-PTFS. CY is supported by  the Max Planck Society in the project MANITOP. AM acknowledges partial support by the Micron Technology Foundation, Inc. AM furthermore acknowledges partial support by the European Union through the FP7 Marie Curie Actions ITN INVISIBLES (PITN-GA-2011-289442) and by the Horizon 2020 research and innovation programme under the Marie Sklodowska-Curie grant agreements No.~690575 (InvisiblesPlus RISE) and No.~674896 (Elusives ITN).
\end{acknowledgments}

\appendix
\section{\label{app:Appendix_RGE}Renormalisation group equations}
We briefly summarise the relevant one-loop RGEs for the scotogenic model, as given in~\cite{Merle:2015ica}. We use a short-hand notation, such that the dependence of couplings on the renormalisation scale $\mu$, is given by $\mu \frac{\mathrm{d}g}{\mathrm{d}\mu} = (4\pi)^{-2} \beta_g$ for any coupling $g$.

The lepton Yukawa RGEs read
\begin{subequations}
\begin{align}
 \beta_{Y_e} &= Y_e \left\lbrace \frac{3}{2}Y_e^\dagger Y_e +\frac{1}{2} h^\dagger h + T - \frac{15}{4} g_1^2 - \frac{9}{4} g_2^2 \right\rbrace, \label{eq:leptonRG}\\
 \beta_{h} &= h \left\lbrace \frac{3}{2} h^\dagger h + \frac{1}{2} Y_e^\dagger Y_e + T_\nu - \frac{3}{4} g_1^2 - \frac{9}{4} g_2^2 \right\rbrace,
 \label{eq:neutrinoRG}
\end{align}
\end{subequations}
where we have abbreviated $T_\nu \equiv \textrm{Tr}\left(h^\dagger h \right)$ and $T \equiv \textrm{Tr}\left(Y_e^\dagger Y_e + 3 Y_u^\dagger Y_u + 3 Y_d^\dagger Y_d\right)$. The right-handed singlet fermion masses obey:
\begin{equation}
 \beta_{M} = \left\lbrace \left(h\, h^\dagger\right) M + M \left(h\, h^\dagger \right)^* \right\rbrace.
 \label{eq:RHnuRG}
\end{equation}
With the short-hand notations $T_{4\nu}\equiv \mathrm{Tr} \left( h^\dag h \, h^\dag h \right)$ , $T_{4}\equiv \mathrm{Tr} \left( Y_e^\dag Y_e Y_e^\dag Y_e + 3 Y_u^\dag Y_u Y_u^\dag Y_u + 3 Y_d^\dag Y_d Y_d^\dag Y_d \right)$, and $T_{\nu e}\equiv \mathrm{Tr} \left( h^\dag h \,Y_e^\dag Y_e \right)$, the RGEs of the quartic self-couplings are:
\begingroup
\setlength{\jot}{8pt}
\begin{subequations} \label{eq:lambdaRGEs}\allowdisplaybreaks
\begin{align} 
 \begin{split} \label{eq:RGlambda1}
     \beta_{\lambda_1} =& 12 \lambda_1^2 + 4 \lambda_3 ( \lambda_3 + \lambda_4 ) + 2 \lambda_4^2  \\
	& + 2 \lambda_5^2 + \frac{3}{4} \left(g_1^4 + 2 g_1^2 g_2^2 + 3 g_2^4\right) \\
	& - 3 \lambda_1 \left(g_1^2 + 3g_2^2\right) + 4 \lambda_1 T - 4 T_4,
 \end{split}\\
 \begin{split} \label{eq:RGlambda2}
  \beta_{\lambda_2} =& 12 \lambda_2^2 +4 \lambda_3 (\lambda_3 + \lambda_4) + 2 \lambda_4^2 \\
	&  + 2 \lambda_5^2 - 3 \lambda_2 \left(g_1^2 + 3g_2^2\right) + 4 \lambda_2 T_\nu  \\
	& + \frac{3}{4} \left(g_1^4 + 2 g_1^2 g_2^2 + 3 g_2^4\right)- 4 T_{4\nu},
 \end{split} \\
 \begin{split} \label{eq:RGlambda3}
  \beta_{\lambda_3} =& 
	2 \left(\lambda_1 + \lambda_2 \right) \left( 3\lambda_3 + \lambda_4\right) + 4\lambda_3^2 \\
	& + 2 \lambda_4^2 + 2 \lambda_5^2 - 3 \lambda_3 \left(g_1^2 + 3g_2^2\right) \\
	&  + \frac{3}{4} \left(g_1^4 - 2 g_1^2 g_2^2 + 3 g_2^4\right) \\
	& + 2 \lambda_3 \left(T +T_\nu\right) - 4 T_{\nu e},
 \end{split} \\
 \begin{split} \label{eq:RGlambda4}
  \beta_{\lambda_4} =& 
	2 \left(\lambda_1 + \lambda_2 \right) \lambda_4 + 8 \lambda_3 \lambda_4 + 4 \lambda_4^2 \\
	& + 8 \lambda_5^2 + 3 g_1^2 g_2^2 - 3 \lambda_4 \left(g_1^2 + 3g_2^2\right) \\
	& + 2 \lambda_4 \left(T +T_\nu\right) + 4 T_{\nu e},
 \end{split} \\
 \begin{split}
    \beta_{\lambda_5} =& \lambda_5 \big[ 2\left(\lambda_1 + \lambda_2\right) + 8\lambda_3 +12\lambda_4 \\
	  & - 3 \left(g_1^2 + 3g_2^2\right) + 2 \left(T + T_\nu\right)\big].  \label{eq:RGlambda5}
  \end{split}
\end{align}
\end{subequations}
\endgroup
Here, it is in fact the term ``$- 4 T_{4\nu}$'' in Eq.~\eqref{eq:RGlambda2} that is truly dangerous for the stability of the vacuum. Finally, the scalar mass parameters have the following RGEs:
\begingroup
\setlength{\jot}{8pt}
\begin{subequations}
\begin{align} 
  \begin{split}\label{eq:m1RG}
    \beta_{m_H^2} =& 6 \lambda_1 m_H^2 +2\left(2\lambda_3 + \lambda_4\right)m_\eta^2 \\
    & + m_H^2\left[ 2T - \frac{3}{2} \left(g_1^2 + 3g_2^2\right)\right],
  \end{split}\\
  \begin{split}\label{eq:m2RG}
    \beta_{m_\eta^2} =& 6 \lambda_2 m_\eta^2 +2\left(2\lambda_3 + \lambda_4\right)m_H^2 \\
    & + m_\eta^2\left[ 2T_\nu - \frac{3}{2} \left(g_1^2 + 3g_2^2\right)\right] \\
    & - 4 \sum_{i=1}^3 M_i^2\left(h \, h^\dagger\right)_{ii}.
  \end{split}
\end{align}
\end{subequations}
\endgroup 


\bibliographystyle{./apsrev}
\bibliography{./literature}

\end{document}